\documentclass[11pt]{article}
\usepackage{moriond,epsfig}

\def\Journal#1#2#3#4{{#1} {\bf #2}, #3 (#4)}

\def\PRL{\em Phys. Rev. Lett.}

\def\be{\begin{equation}}
\def\ee{\end{equation}}
\def\bea{\begin{eqnarray}}
\def\eea{\end{eqnarray}}

\def\snn{$\sqrt{s_{NN}}=$}
\def\v2{$v_{2}$}
\begin{document}
\vspace*{4cm}
\title{THE IMPORTANCE OF THE INITIAL GEOMETRY \\IN HEAVY ION COLLISIONS}

\author{ R.S. HOLLIS$^{6}$ for the PHOBOS Collaboration\\
\vspace{2mm}
B.Alver$^4$,
B.B.Back$^1$,
M.D.Baker$^2$,
M.Ballintijn$^4$,
D.S.Barton$^2$,
R.R.Betts$^6$,
R.Bindel$^7$,
W.Busza$^4$,
Z.Chai$^2$,
V.Chetluru$^6$,
E.Garc\'{\i}a$^6$,
T.Gburek$^3$,
K.Gulbrandsen$^4$,
J.Hamblen$^8$,
I.Harnarine$^6$,
C.Henderson$^4$,
D.J.Hofman$^6$,
R.S.Hollis$^6$,
R.Ho\l y\'{n}ski$^3$,
B.Holzman$^2$,
A.Iordanova$^6$,
J.L.Kane$^4$,
P.Kulinich$^4$,
C.M.Kuo$^5$,
W.Li$^4$,
W.T.Lin$^5$,
C.Loizides$^4$,
S.Manly$^8$,
A.C.Mignerey$^7$,
R.Nouicer$^2$,
A.Olszewski$^3$,
R.Pak$^2$,
C.Reed$^4$,
E.Richardson$^7$,
C.Roland$^4$,
G.Roland$^4$,
J.Sagerer$^6$,
I.Sedykh$^2$,
C.E.Smith$^6$,
M.A.Stankiewicz$^2$,
P.Steinberg$^2$,
G.S.F.Stephans$^4$,
A.Sukhanov$^2$,
A.Szostak$^2$,
M.B.Tonjes$^7$,
A.Trzupek$^3$,
G.J.van~Nieuwenhuizen$^4$,
S.S.Vaurynovich$^4$,
R.Verdier$^4$,
G.I.Veres$^4$,
P.Walters$^8$,
E.Wenger$^4$,
D.Willhelm$^7$,
F.L.H.Wolfs$^8$,
B.Wosiek$^3$,
K.Wo\'{z}niak$^3$,
S.Wyngaardt$^2$,
B.Wys\l ouch$^4$\\
\vspace{3mm}
\small
%
%
%
%
$^1$~Argonne National Laboratory, Argonne, IL 60439-4843, USA\\
$^2$~Brookhaven National Laboratory, Upton, NY 11973-5000, USA\\
$^3$~Institute of Nuclear Physics PAN, Krak\'{o}w, Poland\\
$^4$~Massachusetts Institute of Technology, Cambridge, MA 02139-4307, USA\\
$^5$~National Central University, Chung-Li, Taiwan\\
$^6$~University of Illinois at Chicago, Chicago, IL 60607-7059, USA\\
$^7$~University of Maryland, College Park, MD 20742, USA\\
$^8$~University of Rochester, Rochester, NY 14627, USA\\}

\address{}


\maketitle\abstracts{Elliptic flow, elliptic flow fluctuations and
fluctuations in the initial geometry point to a description of nuclear
collisions that is driven by the initial geometry, a quantity which
appears to be imprinted from the instant of the collision.  In these
proceedings, recent results from the PHOBOS collaboration are discussed
in the context of the importance of the collision geometry.}

\section{Introduction}

Since the start of the RHIC program the measurement of particle
azimuthal anisotropy, or flow, has been considered as one of the most
important probes of nuclear collisions.  Elliptic flow, in particular,
is an important property of particle production as it is sensitive to
the early stages of the collision and thus its study affords unique
insights into the properties of the hot, dense matter that is produced
in these collisions.  At the root of this measurement lies a connection
to the initial overlap geometry of the colliding nuclei, in particular
the eccentricity of the initial overlap region of nucleons, which can be
discussed as an averaged or event-by-event property of the system.  The
PHOBOS experiment has measured the elliptic flow for Au+Au and Cu+Cu
collisions from \snn~19.6 to 200~GeV, versus centrality and transverse
momentum.  For 200~GeV Au+Au collisions, a new analysis of the fluctuations
in the magnitude of elliptic flow have revealed a startling agreement with
a simple geometrical model of nuclear collisions.

\section{Initial Collision Geometry}
\begin{figure}[t]
\hspace{0.05\textwidth}
\psfig{figure=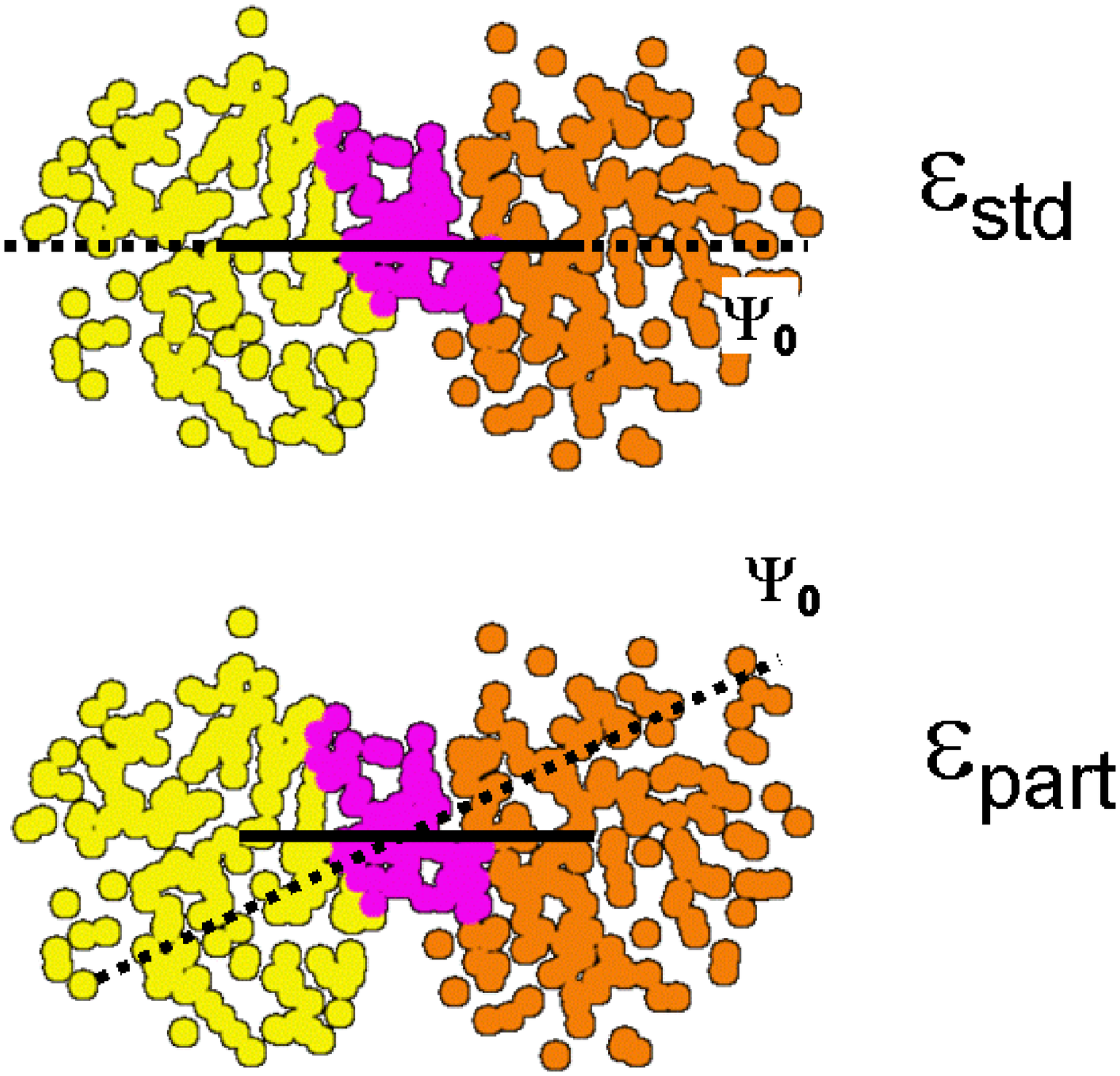,width=0.35\textwidth}
\hspace{0.07\textwidth}
\psfig{figure=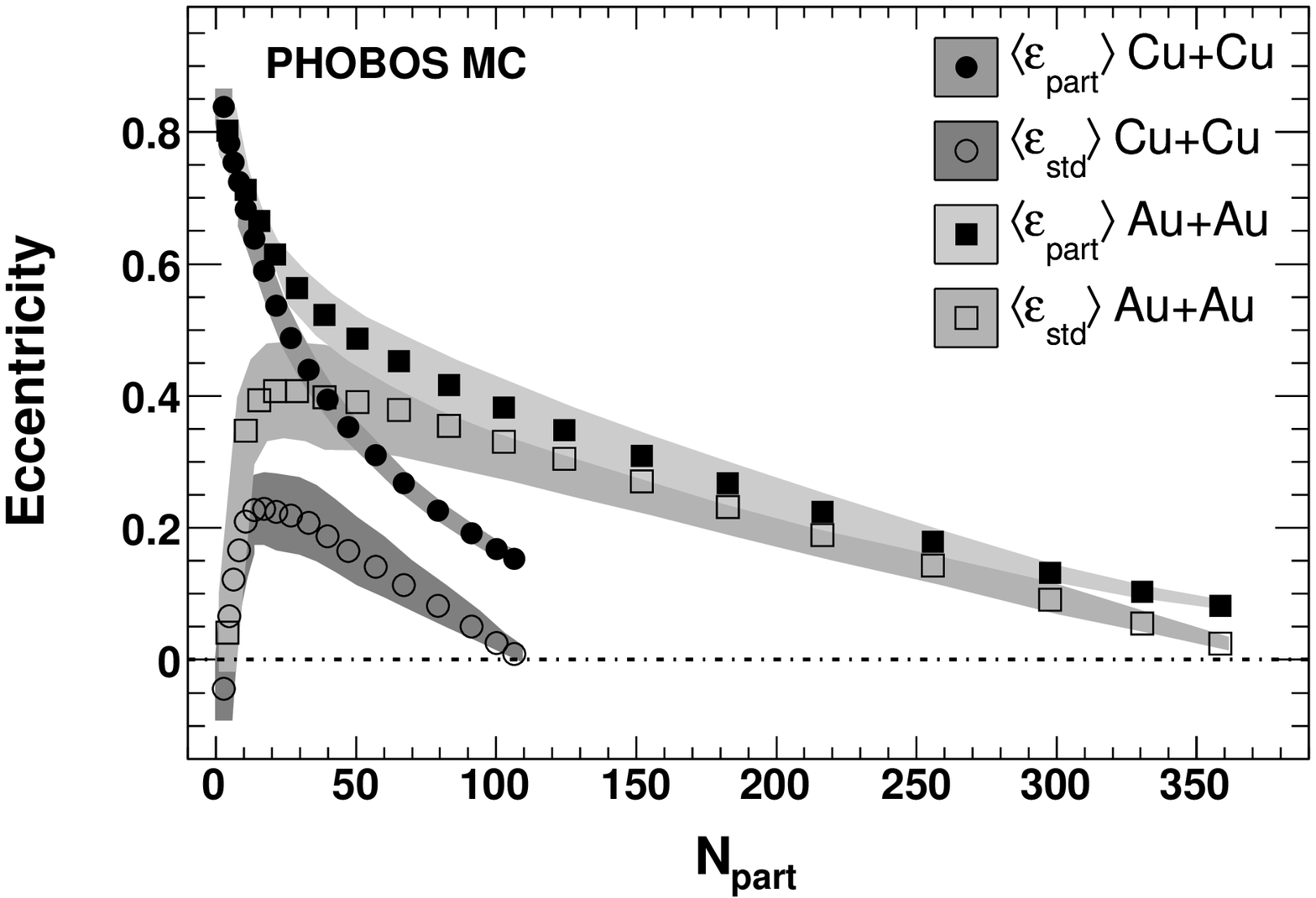,width=0.475\textwidth}
\caption{The left panel visualizes the two approaches to calculating
eccentricity.  The purple region (at center) in each collision illustrates
the interacting nucleons.  The orange and yellow nucleons (away from
collision zone) are assumed not to directly influence the eccentricity.
The solid (dashed) line represents the collision (participant) reaction
plane.  The lower part shows that the assumed reaction plane is rotated
into the plane which maximizes the eccentricity, i.e. aligned along the
semi-minor axis of the participant region.  The right panel shows the
difference of these two approaches for both Au+Au and Cu+Cu collisions.
Cu+Cu collisions show a significant difference in the calculated
eccentricity, whilst the discrepancy is less for Au+Au collisions.
\label{fig:PHOBOS_EccDefs}}
\end{figure}

The collision geometry has always played an important role in heavy-ion
collision analysis.  The most simplistic definitions of centrality, 
derived from a Glauber model~\cite{cite:Glauber}, and consequently the
number of nucleons, $N_{part}$, expected to have participated in the
collision is fundamental to this area of high-energy physics.  As well
as $N_{part}$, additional information can be gained from this model,
including the spatial anisotropy of the collection of participating
nucleons, or eccentricity ($\epsilon$).  This anisotropy leads to the
observed elliptic flow signal in data, discussed in the next sections.
There are several methods for calculating $\epsilon$, two of which are
illustrated in Fig.~\ref{fig:PHOBOS_EccDefs}.  On the left, a schematic
depiction of the ``standard'' (top, $\epsilon_{std}$) and ``participant''
(bottom, $\epsilon_{part}$) methods are shown.  The former assumes that
the collection of participating nucleons is oriented such that the
semi-minor axis is aligned along the {\it reaction plane} - through the
centers of the original colliding nuclei.  As one can see, this is not
always the case and may thus result in a reduced eccentricity. For the
participant method, the semi-minor axis is allowed to rotate, such that
the eccentricity is maximized.  Eqn.~\ref{eqn:eccstd} is a mathematical
representation of the eccentricity for both methods.

\begin{equation}
\epsilon_{std}=\frac{\sigma^{2}_{y}-\sigma^{2}_{x}}{\sigma^{2}_{y}+\sigma^{2}_{x}}
\label{eqn:eccstd}
\hspace{0.25\textwidth}
\epsilon_{part}=\frac{\sqrt{(\sigma^{2}_{y}-\sigma^{2}_{x})^{2}+4\sigma^{2}_{xy}}}{\sigma^{2}_{y}+\sigma^{2}_{x}}
\label{eqn:eccpart}
\end{equation}

The difference in mean eccentricity between these two methods can be seen
on the right panel of Fig.~\ref{fig:PHOBOS_EccDefs}.  For central Au+Au
collisions little difference is observed between the two.  For more
peripheral Au+Au or Cu+Cu collisions, large differences are seen, due 
primarily to the finite number of participating nucleon in such collisions.
This difference in the magnitude of the eccentricity calculated using both
methods from the model is observed in the elliptic flow data.

\begin{figure}[t]
\psfig{figure=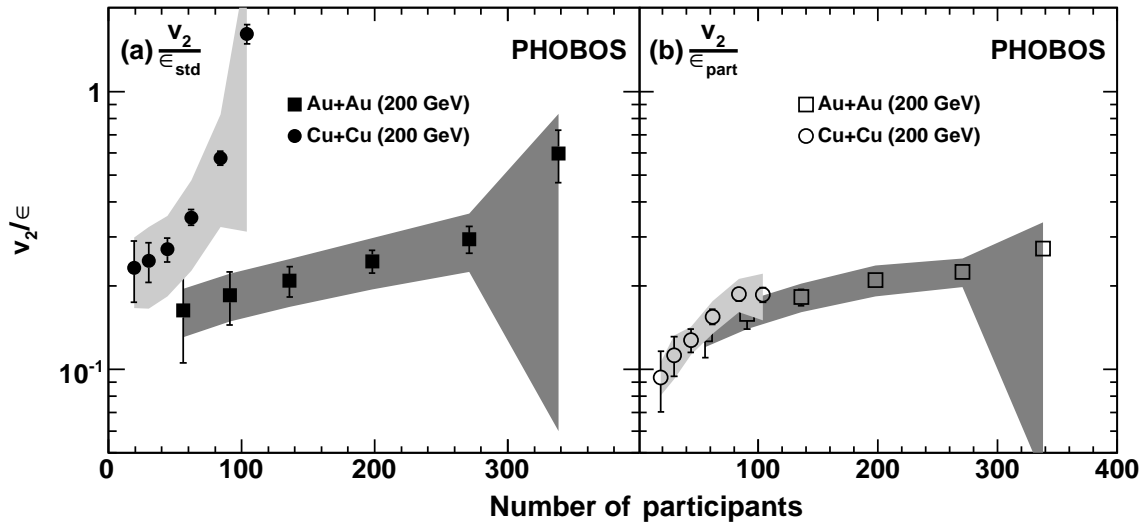,width=1.10\textwidth}
\caption{The elliptic flow, v$_{2}$, scaled by the eccentricity from
a Glauber model calculation for the (a) standard and (b) participant
approaches.  Data are for Au+Au and Cu+Cu collisions at $\sqrt{s_{NN}}=200$GeV.
Shaded bands (error bars) represent the systematic (statistical) uncertainty.
\label{fig:PHOBOS_v2_Ecc}}
\end{figure}

\section{Elliptic Flow}
Measurements of the elliptic flow, \v2, from PHOBOS are made over a
broad range of pseudorapidity, centrality and energy.  Generic features
of particle production are found for both the Au+Au and Cu+Cu systems.
At midrapidity, for similar centrality selections, the magnitude of \v2
increases from the lowest collision energy of \snn~19.6~GeV up to
200~GeV~\cite{cite:PHOBOS_Flow_Energy}.  The magnitude of the \v2 diminishes
as the pseudorapidity increases (for more forward particles) and is found to
have a roughly triangular shape~\cite{cite:PHOBOS_Flow_Energy}.  The coupling
of the collision energy and pseudorapidity dependences result in the \v2
signal exhibiting an extended longitudinal scaling
behaviour~\cite{cite:PHOBOS_Flow_Energy} whereby the magnitude of \v2 is the
same at the same pseudorapidity relative to beam rapidity (i.e. in the
rest frame of one of the incoming nuclei).

The centrality dependence of \v2 shows the first clear dependence of the
particle distributions following the underlying geometrical
shape~\cite{cite:PHOBOS_Flow_Species}.  For central Au+Au collisions with
an almost full overlap (small impact parameter) both the \v2 and the
eccentricity are found to be small, see Fig.~\ref{fig:PHOBOS_EccDefs}.
As the impact parameter increases, collisions assume an almond shape, and
\v2 and the eccentricity both increase.

For Au+Au collisions, it is found that \v2 scales reasonably with the
standard eccentricity, $\epsilon_{std}$, whereas the Cu+Cu data strongly 
violate this approximate scaling, see Fig.~\ref{fig:PHOBOS_v2_Ecc}a.
Considering the alternate technique, the participant eccentricity, yields
a unification of the two data samples, Fig.~\ref{fig:PHOBOS_v2_Ecc}b.

\section{Elliptic Flow Fluctuations}
The collision species dependence of the integrated elliptic flow signal
is found to be strongly dependent on the collision geometry, and to its
precise definition.  Specifically, the fluctuations in the nucleon
positions on an event-by-event basis appears to drive the final \v2 signal.
If such fluctuations influence the averaged signal, then this should be a
measurable quantity in itself.  One of the latest results from the PHOBOS
collaboration concentrates on measuring these elliptic flow fluctuations.
The method utilizes the whole pseudorapidity coverage of the PHOBOS detector
to measure the \v2 signal on an event-by-event basis, assuming the shape is
either a triangle or a trapeziod.  Details of the analysis method can
be found in Ref.~\cite{cite:PHOBOS_FlowFluc_Tech}.

\begin{figure}
\centering{\psfig{figure=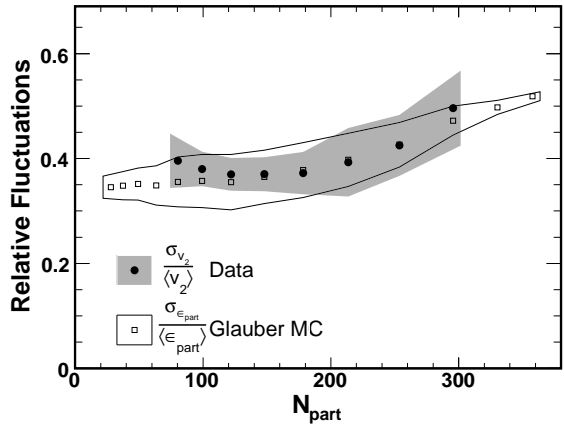,width=0.475\textwidth}}
\caption{Comparison of the elliptic flow fluctuations, $\sigma_{\rm{v}_{2}}$, to
fluctuations in the initial geometry (eccentricity) from a Glauber model.
Data are for Au+Au collisions at $\sqrt{s_{NN}}=200$GeV.
\label{fig:PHOBOS_v2fluc}}
\end{figure}

The elliptic flow fluctuations, expressed as $\sigma_{v_{2}}/v_{2}$, are
shown in Fig.~\ref{fig:PHOBOS_v2fluc}.  The fluctuations are found to be
significant for all centrality classes studied, with a peak close to 50\%
relative fluctuations.  Fluctuations in the eccentricity from the
Glauber model calculations are also found to be significant, with the
magnitude in remarkable agreement with the \v2 fluctuations.  Such an
agreement hints that the detailed initial geometrical configuration is 
imprinted on the final distribution of particles.

\section{Summary}
The initial geometry in nuclear collisions plays an important role in 
particle production at RHIC.  The detailed eccentricity, calculated from
the positions of the interacting nucleons in a Glauber model, has been
shown to unify elliptic flow data from Au+Au and Cu+Cu collisions. The
magnitude of elliptic flow fluctuations are measured and are found to be
large for all centralities.  The level of these fluctuations is strikingly
similar to those from the eccentricity calculations, indicating that the
initial geometry is imprinted on the final particle distributions.

\section*{Acknowledgments}
This work was partially supported by U.S. DOE grants 
DE-AC02-98CH10886,
DE-FG02-93ER40802, 
DE-FG02-94ER40818,  
DE-FG02-94ER40865, 
DE-FG02-99ER41099, and
DE-AC02-06CH11357, by U.S. 
NSF grants 9603486, 
0072204,            
and 0245011,        
by Polish KBN grant 1-P03B-062-27(2004-2007),
by NSC of Taiwan Contract NSC 89-2112-M-008-024, and
by Hungarian OTKA grant (F 049823).

\end{document}